# Recent progress on the fabrication of silver nanowire-based transparent electrodes

Renyun Zhang[1],* and Magnus Engholm[2]
Department of Natural Sciences, Mid Sweden University
Department of Electronics Design, Mid Sweden University
SE-85170 Sundsvall, Sweden

*Email: renyun.zhang@miun.se

## Abstract

Transparent electrodes (TEs) made of metallic nanowires, such as Ag, Au, Cu, and Ni, have attracted rising attention for several reasons: 1) they can act as a substitute for tin oxide-based TEs such as indium-tin oxide (ITO) and fluorine-doped tin oxide (FTO); 2) various methods exist for fabricating such TEs such as filtration, spraying and meyer bar coating; 3) greater compatibility with different substrates can be achieved due to the variety of fabrication methods; and 4) extra functions in addition to serving as electrodes, such as catalytic abilities, can be obtained due to the metals that compose the TEs. There are a large number of applications for TEs, ranging from electronics and sensors to biomedical devices. This short review is a summary of recent progress, mainly during the past five years, on silver nanowire-based TEs. The focus of the review will be on theory development, mechanical, chemical and thermal stability and optical properties. The many applications of TEs are outside the scope of this review.

## 1. Introduction

Transparent electrodes (TEs) are essential components in many optoelectronic applications such as solar cells, touch screen displays and film heaters [1]. The materials used for making these electrodes are mainly transparent conductive oxides (TCOs) such as indium-tin oxide (ITO), fluorine-doped tin oxide (FTO) and doped zinc oxide. However, the low availability of these elements and the high demands imposed by the fabrication conditions lead to high prices for these TEs. In addition, the brittleness of the materials limit their application in new types of flexible electronics such as wearable sensors.

Therefore, new materials, such as conductive polymers, carbon nanotubes, graphene, ultra-thin metal films and metallic nanowires, have been adopted for fabricating TEs. Despite the wide choice of materials, few of these can fulfill the requirements of industrial production (transmittance > 90% and sheet resistance < 100 Ω/sq) [2]. Research has shown that metallic nanowires can fulfill these requirements. Thus, increasing efforts have been made in this area, such as developing new methods for synthesizing nanowires, new deposition methods and new post procedures.

Ag nanowires are the most studied material for making metallic nanowire-based TEs. The most common route is to synthesize Ag nanowires followed by deposition on different substrates using methods such as electrostatic spraying [3,4], doctor blading [5], electroless deposition [6], meyer bar coating [7], ink jet printing [8] and spin coating [9].

Results from studies before 2013 have shown that bare Ag nanowires can easily be deposited on different substrates by using the techniques mentioned above, and the length to diameter ratio of the nanowire determines whether the nanowires have the potential to achieve the requirements for industrial production (transmittance > 90% and sheet resistance < 100 Ω/sq) [2].

Recent trends have been to enhance the flexibility and chemical and thermal stabilities, to tune the optical properties and to apply these TEs in different optoelectronic applications.

This short review will summarize the main advances in silver nanowire-based TEs that have been made in the past five years. The review focuses on theory development, mechanical and chemical stability and tuning of the optical properties.

## 2. Theoretical approaches

### 2.1. Transparency and conductivity

Transparency and conductivity are the two most important parameters for characterizing TEs. For a bulk-like film such as ITO, the transparency, which is usually quantitatively expressed as the transmittance ($T$), is related to the sheet resistance ($R_s$) of the film [10]:

$$T = e^{-\alpha/\sigma_{DC,B} R_s} \qquad (1)$$

where α is the absorption coefficient and $\sigma_{DC,B}$ is the bulk DC conductivity of the film.

Metallic nanowire-based TEs have different mechanisms for their transmittance, based on the free space [10] among the nanowires instead of the absorption coefficient of the material. This means that the density of the nanowires determines the transparency of the TEs. Depending on the density of the nanowires, the transparent films, represented by the relation between the transmittance and the sheet resistance, are found to be either bulk-like or percolative.

In the bulk-like regime, the transmittance can be expressed as:

$$T = \left(1 + \frac{Z_0}{2}\sigma_{Op} t\right)^{-2} \quad (2)$$

where $\sigma_{Op}$ is the optical conductivity ($\sigma_{Op} \approx \alpha/Z_0$) and $Z_0$ is the impedance of free space (377 Ω) [11]. In percolation theory, the conductivity ($\sigma_{DC}$) of the nanowire film is non-linearly related to the difference between the density of the nanowires per unit area, $n$, and the critical density, $n_c$:

$$\sigma_{DC} \propto (n - n_c)^m \quad (3)$$

$n_c$, also called the percolation threshold, [12] is the value at which the network has a percolation probability of ½, and the exponent $m$ has been found, by Monte Carlo simulation, to be 4/3 [13]. The value of $n_c$ is determined by the length of the nanowires ($L_{NW}$) using the equation [16]:

$$n_c = 5.63726/L_{NW}^2 \quad (4)$$

By combining equations 2 and 3, one can predict the relation between the transmittance and the sheet resistance using an equation suggested by Grüner and co-workers [14,15]:

$$T = \left(1 + \frac{Z_0}{2R_s}\frac{\sigma_{Op}}{\sigma_{DC}}\right)^{-2} \quad (5)$$

The experimental data from different materials with transmittance values from 0 to critical values, depending on the structure of the materials, can be fit using equation 4 (Fig. 1). Beyond the critical values, the data deviate from the fitting curves, and percolation theory should be applied.

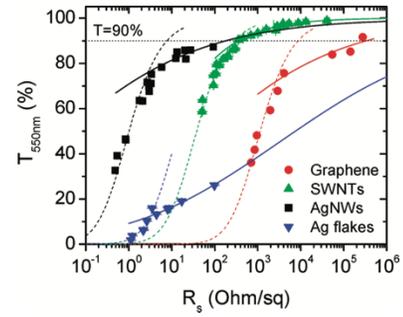

**Fig. 1**. Transmittance (550 nm) plotted as a function of sheet resistance for thin films prepared from four nanostructured materials: graphene, single-walled carbon nanotubes, silver nanowires and silver flakes. The dashed lines represent fits to the bulk regime using eq 2, while the solid lines represent fits to the percolative regime using eq 9. Reprinted with permission from (De, S.; King, P. J.; Lyons, P. E.; Khan, U.; Coleman, J. N. *ACS Nano* **2010**, *4* (12), 7064–7072). Copyright (2010) American Chemical Society

The conductivity, $\sigma_{DC}$, of a nanowire network film can also be expressed using the thickness of the films [13]:

$$\sigma_{DC} \propto (t - t_c)^n \quad (6)$$

where $t_c$ is the threshold thickness and $n$ is the percolation exponent. For a network with a conductivity high enough for industrial production, $t$ is always greater than $t_c$ [10]. Based on this, Coleman and co-workers have further defined the relation between $\sigma_{DC}$ and $\sigma_{DC,B}$ using the equation:

$$\sigma_{DC} = \sigma_{DC,B}\left(\frac{t}{t_{min}}\right)^n \quad (7)$$

where $t_{min}$ is the thickness of the nanowire network film that has $\sigma_{DC}$ equal to $\sigma_{DC,B}$, usually 2.33 times the diameter of the nanowires [10].

Coleman et al. [10] have further used $n$ to define a percolation Figure of Merit, $\Pi$, that defines the relation between the transmittance $T$ and the sheet resistance $R_S$ of a TE:

$$\Pi = 2\left[\frac{\frac{\sigma_{DC,B}}{\sigma_{Op}}}{(Z_0 t_{min}\sigma_{Op})^n}\right]^{\frac{1}{n+1}} \quad (8)$$

$$T = \left[1 + \frac{1}{\Pi}\left(\frac{Z_0}{R_S}\right)^{\frac{1}{n+1}}\right]^{-2} \quad (9)$$

### 2.2. Transmission and film parameters

Mutiso and co-workers have developed another method to simulate nanowire-based networks. In their work, the sheet resistance is calculated based on the effective contact resistance ($R_{c\_effective}$) between the nanowires. They use the area fraction ($AF$) to define the density of the nanowires and, subsequently, the transmittance using the empirical equation [16]:

$$\%T = 100 - a_1 AF \quad (10)$$

where $a_1$ is a fitting parameter that accounts for the diameter and wavelength-dependent optical properties of the nanowires.

A plot of the transmittance and the sheet resistance shows a perfect fit using this method in both the bulk-like and percolative regimes. However, the ratio between the lengths and diameters must be given in order to perform the simulation.

Recently, Khanarian and co-workers [18] have further developed the Mie light scattering theory of spheres and applied it to nanowires to predict both the transmission and the haze.

The transmission of a Ag-nanowire film with thickness d is given by:

$$T = e^{-n_v C_{ext} d} \quad (11)$$

where $C_{ext}$ is the extinction coefficient of the nanowires that is given by Mie [17] and $n_v$ is the number of nanowires per volume, which is related to the volume fraction $\emptyset_v$, the diameter $D$, and the length $L$ by the equation $n_v = \frac{\emptyset_v}{L\pi\left(\frac{D^2}{4}\right)}$.

For TEs made of Ag nanowires with a thickness less than 100 nm, the transmission has been found to be better represented by:

$$T = e^{-n_s C_{ext}} \quad (12)$$

where $n_s$ is the number of nanowires per unit area.

### 2.3. Haze

Haze refers to the amount of light that is subject to wide scattering when passing through nanowire films. It is an important parameter for describing the optical properties of nanowire-based TEs. However, theoretical approaches that predict haze are limited.

Khanarian and co-workers [18] have extended the derivation provided by Willmouth [19] to derive an expression for haze in Ag nanowire films. The geometry of the scattered light passing through a Ag nanowire film is schematically presented in Fig. 2. They further assumed that the Fresnel transmission terms [21] could be approximated by their average values at each optical interface. Then, an expression to predict the haze for nanowires was derived as follows:

$$H = \frac{F_{12} F_{23} \frac{\langle C_{2.5}^{90}\rangle}{C_{ext}}(1-e^{-\phi_s \langle Q_{ext}\rangle})}{F_{12} F_{23} F_{31} e^{-\phi_s \langle Q_{ext}\rangle} + F_{12} F_{23} \frac{\langle C_0^{90}\rangle}{C_{ext}}(1-e^{-\phi_s \langle Q_{ext}\rangle})} \quad (13)$$

where $F_{ij}$ are the Fresnel transmission terms, $Q_{ext}$ is the scattering efficiency, $C_m^n$ is the scattering efficiency between angles $m$ and $n$, and $\emptyset_s$ is the surface fraction of the nanowires.

It is possible to predict the relation between the diameter of the nanowires and the ratio $H(\%)/\emptyset_s$. Using this relation, it is easy to estimate the surface fraction of the nanowires based on their diameters (Fig. 3). However, this relation is based on the assumption of no overlap effects or multiple scattering events among the nanowires and needs to be adjusted for high $\emptyset_s$.

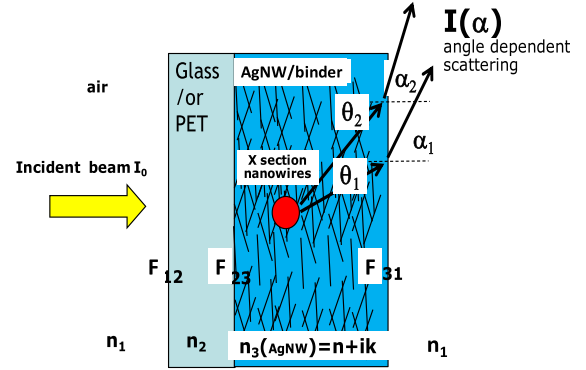

**Fig. 2.** Scattering geometry of nanowires in films. Reprinted from [Khanarian, G.; Joo, J.; Liu, X.-Q.; Eastman, P.; Werner, D.; O'Connell, K.; Trefonas, P. *J. Appl. Phys.* **2013**, *114* (2), 24302.:10.1063/1.4812390.] with the permission of AIP Publishing.

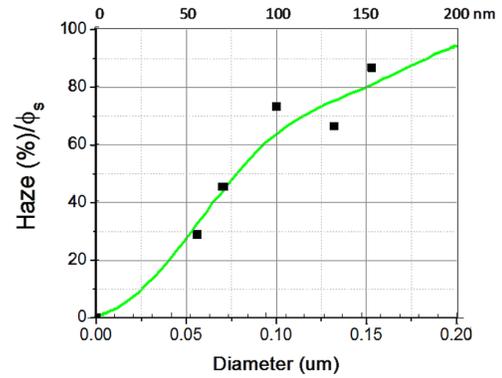

**Fig. 3.** Slope of haze (%) versus surface fraction $\emptyset_s$, Haze $(\%)/\emptyset_s$, as a function of the Ag nanowire diameter. Reprinted from [Khanarian, G.; Joo, J.; Liu, X.-Q.; Eastman, P.; Werner, D.; O'Connell, K.; Trefonas, P. *J. Appl. Phys.* **2013**, *114* (2), 24302. DOI:10.1063/1.4812390.] with the permission of AIP Publishing.

It follows from Fig.3 that the haze is strongly dependent on the diameter of the Ag nanowires. Hence, for applications that require a low haze (e.g. displays, H < 1%) that is found possible only for diameters less than ~50 nm.

Preston et al.[20] performed simulations and experiments on Ag nanowires and also came to the same conclustion. TEs with smaller Ag nanowire diamters have a lower haze and exhibit a better performance.

## 3. Experimental approaches

### 3.1. Mechanical stability
#### 3.1.1. Flexibility

Flexibility is a crucial factor for TEs, as it is of great importance for their application in soft electronics, e.g., as wearable sensors. Flexibility is a factor that combines the mechanical and electrical stability of a TE and is usually characterized by plotting the electrical conductivity versus the cycles of mechanical bending. Several protocols have been developed for enhancing the mechanical flexibility of TEs, such as welding or the deposition of extra material.

Welding is one of the more effective ways to enhance the flexibility of TEs while retaining their electrical properties. Thermal annealing, light irradiation, mechanical pressing, plasma treatment, extra coating, cold welding and chemical treatment are methods that have been used for welding Ag nanowires.

Hwang and co-workers have demonstrated the welding of Ag nanowires by post-annealing at 180 °C for 25 min [4], thereby forming fused-in junctions among the nanowires. This welding procedure can significantly enhance the flexibility of TEs, while retaining their electrical conductivity after 500,000 cycles with a strain of 1 %.

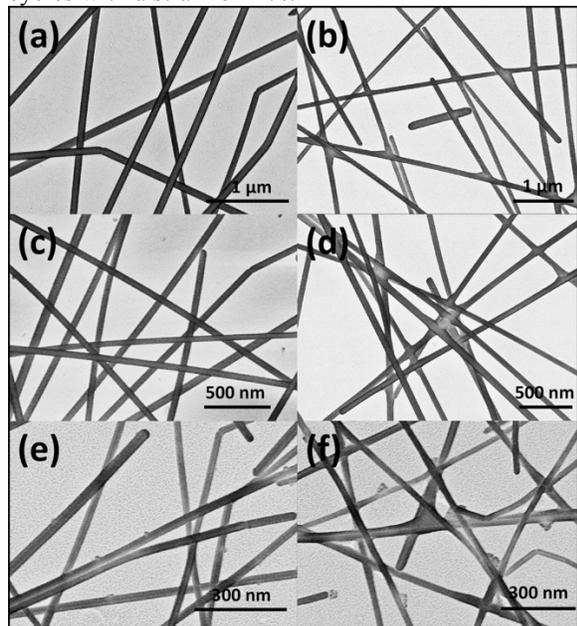

**Fig. 4.** SEM images of transparent electrodes fabricated from a,b) 100, c,d) 60, and e,f) 20 nm silver nanowires before and after chemical treatment. From (Lu, H.; Zhang, D.; Cheng, J.; Liu, J.; Mao, J.; Choy, W. C. H. *Adv. Funct. Mater.* **2015**. DOI:10.1002/adfm.201501004). Copyright © 2015 by John Wiley Sons, Inc. Reprinted by permission of John Wiley & Sons, Inc.

The welding of Ag nanowires with intense light is also an effective method for enhancing the flexibility of TEs. Kou et al. [1] have reported the welding of Ag nanowires using simulated sunlight (0.1 W/cm$^2$). The welded TEs were shown to achieve the same flexibility as those welded at 200 °C. A 1.8 x 1.8 cm sized TE has been found to maintain its electrical conductivity after 500 cycles of bending to a minimum radius of curvature of <0.15 cm. In addition to simulated sunlight, flashlight irradiation has also been used to weld Ag nanowires. Li et al. [21] used flashlight irradiation with an energy of 4.6 J/cm$^2$ per pulse to weld Ag nanowires on paper substrates. Hwan and co-workers [22] have used a higher power density of 10.3 J/cm$^2$ to weld Ag nanowires with an average diameter of 35 nm. High power lamps have also been used to weld Ag nanowires [23], where 10 to 120 seconds of irradiation under a tungsten-halogen lamp with a power density of 30 W/cm$^2$ can weld the junctions between the nanowires.

Plasma treatment has also been found to cause the welding of Ag nanowires. Zhu and co-workers treated Ag nanowire films with 75 W plasma [24] at room temperature. The treated film maintained its electrical conductivity after 10,000 bending cycles at a frequency of 2 Hz.

Welding can also be done by post-deposition of 50 nm fluorine-doped ZnO (FZO) using a laser deposition method [9]. Using this method, no change in electrical conductivity was observed after 1000 bending cycles, although the strain of the bending was not reported. Cheong et al. have [25] welded Ag nanowires by depositing 30 and 50 nm indium tin oxide (ITO) using a DC magnetron sputtering system. The welded films maintained their electrical conductivities during 10,000 bending cycles.

Capillary-force-induced cold welding [26] is a new method to create self-limited welding of the wire-wire junctions. These processes can be performed simply by applying moisture to the Ag nanowire films. However, this cold welding has not been fully studied, and the electrical conductivity decreased at a bending radius of 1.5 mm.

Chemical methods have also been reported to weld Ag nanowire films [27]. These methods are used for the deposition of Ag atoms at the junctions of the nanowires (Fig. 4) because of the high local chemical potential of the concave surface [28]. The process can be easily performed by dipping the films in silver-containing solutions such as silver-ammonia [29].

### 3.1.2. Adhesion

Adhesion is another important parameter for the mechanical stability of TEs. The Ag nanowire films are not self-supporting, and thus, supporting substrates are required to hold the films. However, bare Ag nanowire films do not adhere well to most substrates, and additional procedures are required to enhance the adhesion.

There are two general routes for enhancing the adhesion between Ag nanowires and substrates. The first one directly increases the adhesion by using extra processes; the second one requires the addition of other materials.

Spontaneous heating of the Ag nanowires and the substrate can lead to stronger contact between the nanowires and the substrate, resulting in stronger adhesion. Lee and co-workers [30] have laminated Ag nanowires on a PET substrate at 120 °C to enhance the adhesion. Alternatively, other research groups have used intense-pulse-light (IPL) to increase the adhesion between the nanowires and a substrate [31,32]. The IPL method can heat the nanowires and the substrate to a high temperature for a short period, thus increasing the contact between the materials.

In addition to heating, forces added to the films can increase the adhesion. A strong conformal pressure applied to the Ag nanowires can improve their adhesion to PET [33]. Liu and co-workers have noted that the capillary force produced by moisture on a Ag nanowire film can pull the nanowires to the substrate [26].

Although the methods mentioned above have been shown to increase the adhesion, they are less popular than those assisted by extra coating or by compositing the nanowire with substrate materials.

The coating of an extra layer, such as ZnO [34], aluminum-doped ZnO (AZO) [35] or TiO$_2$ [36], by either ALD or sol-gel methods has been found to significantly increase the adhesion. Graphene has also been observed to increase the adhesion of Ag nanowires to a polyester yarn surface [37]. This is due to the existence of a large van der Waals force between graphene and the polyester (Fig. 5), which helps to bind the Ag nanowires.

Compositing Ag nanowires with substrate polymers is another way to increase the adhesion. Nam and co-workers [38] have embedded Ag nanowires in Norland Optical Adhesive (NOA) 63 to obtain strong adhesion. Other materials, such as chitosan [39], alginate [40], and polyvinyl alcohol [41], have also been composited with Ag nanowires, resulting in improved adhesion.

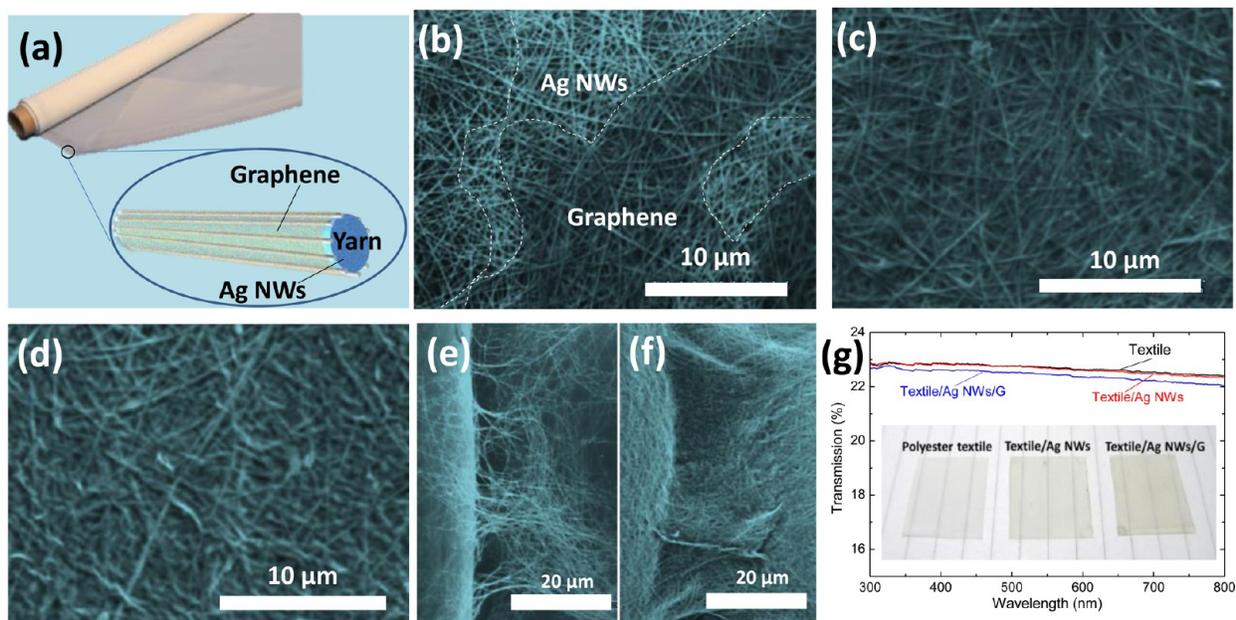

**Fig. 5.** (a) Schematic diagram of an e-textile with a polyester/Ag NW/graphene core−shell structure. (b−d) SEM images of polyester/Ag NW/graphene samples with different numbers of graphene-coating cycles: (b) one cycle, (c) two cycles, and (d) three cycles. SEM images of the fiber cross-linked regions (e) before graphene coating and (f) after graphene coating. (g) Visible-light transmittance of the textile, the textile/Ag NW, and the textile/Ag NW/graphene samples; the insets are photographs. Reprinted with permission from (Wu, C.; Kim, T. W.; Li, F.; Guo, T. *ACS Nano* **2016**, *10* (7), 6449–6457. DOI:10.1021/acsnano.5b08137). Copyright (2016) American Chemical Society.

### 3.1.3. Stretchability

Stretchability is specifically required for Ag nanowire TEs that need to withstand tensile, compressive and shear forces. To meet this requirement, both the Ag nanowire film and the substrate should be stretchable.

Hydroxylated polydimethylsiloxane (PDMS) [42–44] and polyurethane urea (PUU) [45] are two of the most commonly used substrate materials for stretchable TEs due to their excellent elasticity. In addition, poly(urethane acrylate) (PUA) [46] and polyimide [31] have also been used for supporting Ag nanowire layers. There are two main routes for depositing Ag nanowires on these substrates. One is to prepare the Ag nanowire film first and to then transfer it onto the substrate [42–44]; the other is to fabricate the Ag nanowire-polymer composite first and to then form a film of the composite [46].

The elasticity of the Ag nanowire films is the core of the stretchable TEs and determines the performance of the TEs. Since the Ag nanowires themselves are not elastic, structural designs are needed to ensure the electrical properties of the film under strain.

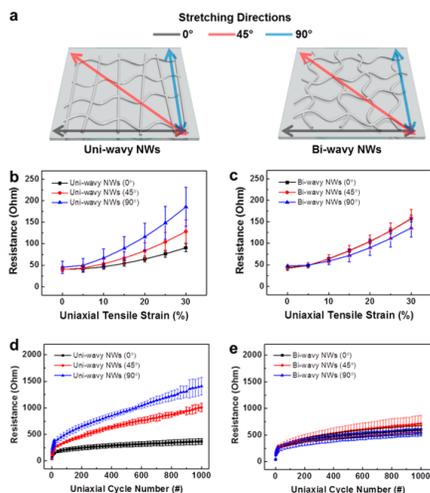

**Fig. 6.** (a) Schematics showing the stretching directions of the uniaxial tensile strains for uniwavy NWs and biwavy NWs. Resistance changes as a function of the applied uniaxial tensile strain for (b) uniwavy NWs and (c) biwavy NWs. Resistance changes as a function of the uniaxial cyclic strain for (d) uniwavy NWs and (e) biwavy NWs. Electrodes were repeatedly stretched to 30% and released back to 0%. Reprinted with permission from (Kim, K. K.; Hong, S.; Cho, H. M.; Lee, J.; Suh, Y. D.; Ham, J.; Ko, S. H. *Nano Lett.* **2015**, *15* (8), 5240–5247. DOI:10.1021/acs.nanolett.5b01505). Copyright (2015) American Chemical Society.

A random network of Ag nanowires can lead to decreased electrical conductivity when multiple cycles of mechanical strain are applied [46]. However, if the Ag nanowires are pre-strained to form a wrinkled structured film (Fig. 6), the stretchability of the TEs is significantly enhanced [43]. Another protocol, reported by Kim and co-workers, is to make Ag TEs with wavy structured Ag nanowires. This was done by compressing the floating Ag nanowire film before it was transferred onto PDMS [42].

In addition to pre-processing the Ag nanowire films, one can also make more complex structures to enhance the performance. Kim and co-workers have patterned polyepoxy acrylate (PEA) islands on Ag nanowires deposited on PUA substrates (Fig. 7). Using this approach, the stretchability increased significantly, and a 60 % strain was sustained for 500 cycles [31].

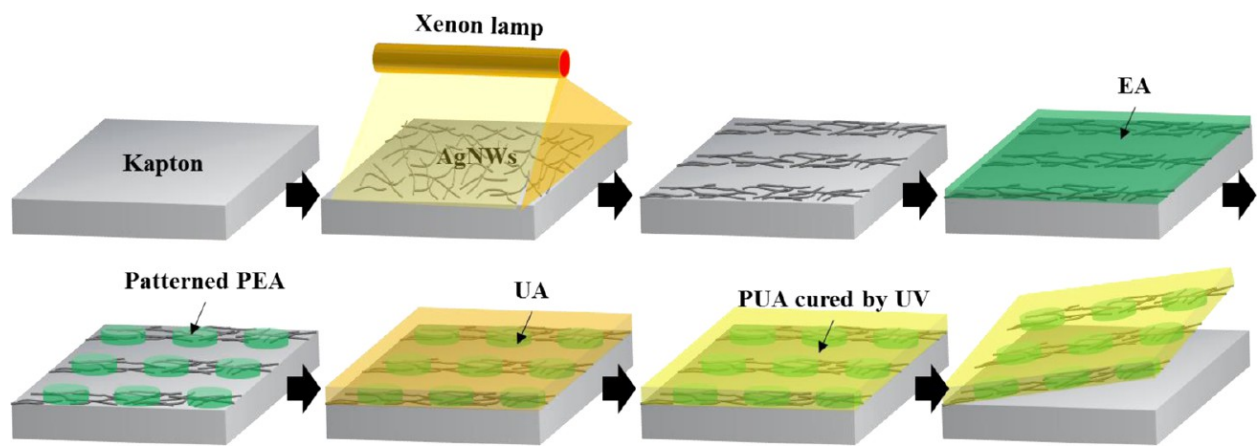

**Fig. 7.** Scheme showing the fabrication of a heterogeneous Ag NW/polymer composite structure. Reprinted with permission from (Kim, Y.; Jun, S.; Ju, B.-K.; Kim, J.-W. *ACS Appl. Mater. Interfaces* **2017**, *9* (8), 7505–7514. DOI:10.1021/acsami.6b11853). Copyright (2017) American Chemical Society.

### 3.2. Chemical stability

Since silver can react with oxygen and acids, a Ag nanowire-based TE is not chemically stable when it is exposed to air and acids. Additional procedures must be used to enhance the chemical stability.

The most common way to protect Ag nanowires from chemical reactions is to encapsulate or polymerize the nanowires with a layer of inorganic or organic material. This layer will prevent direct contact between the Ag nanowires and potential reactants.

Metal oxides are the most commonly used inorganic materials for the protection of Ag nanowires from oxidation. Atomic layer deposition (ALD) of $Al_2O_3$ onto Ag nanowires has been found to effectively protect the nanowires from oxidation [5,47,48]. Alternatively, ZnO [35] and $TiO_2$ [30] deposited by ALD have also been shown to have the same protective effect. Despite their anti-oxidation effects, some of the metal oxide coatings, such as ZnO and $Al_2O_3$, are not chemically stable in acidic conditions, thereby limiting their application in certain environments.

Glass fabric materials have been used to enhance the chemical stability of Ag nanowire films. A film with a glass-fabric-reinforced transparent composite (GFRHybrimer) embedded with the Ag nanowire networks showed more resistance to corrosive reagents such as 5 wt% $K_2S$ [49].

Polymers are alternative materials for protecting Ag nanowires through encapsulation [39] or polymerization [50,51]. In many cases, the polymers are added for other purposes such as enhancing the adhesion [52]. However, the protective effects are spontaneously added to the Ag nanowires.

### 3.3. Thermal stability

Thermal stability is required for the application of TEs in high temperature environments. The thermal stability of the polymer substrates is also crucial to TEs, but this issue will not be considered in this review. Instead, this section will focus on the thermal stability of the Ag nanowire film.

Ag nanowires are thermally stable at temperatures below 200 °C, while higher temperatures lead to decomposition [53]. The most common phenomenon observed after heating Ag nanowires is that the nanowires shrink upon heating and form droplets. This phenomenon is also elsewhere called negative thermal expansion [54]. This phenomenon is caused by the coalescence of Ag nanowires [55]. This coalescence is due to the high surface energy of the nanowires and is size [56] and temperature [55] dependent. Smaller nanostructures have higher surface energies [56] and are thus more likely to coalesce, while higher temperatures accelerate the coalescence process [57]. The result of the coalescence is that the nanowires break into small droplets. The origin of the droplet-forming behavior of heated nanowires is called Rayleigh-Plateau instability [12].

Surface coating with metal oxides is a common way to reduce the Rayleigh-Plateau instability and thus enhance the thermal stability. A Ag nanowire film that was coated using the ALD method by ZnO with a thickness of 4.5 nm had long-term thermal stability up to 300 °C [47]. The stability can be improved up to 350 °C by coating with 40 nm thick ZnO [53]. Greater stability was also found for $Al_2O_3$-coated Ag nanowire films (Fig. 8), where a 5.3 nm thick $Al_2O_3$ layer increased the stability to 380 °C [5]. $TiO_2$ is another material that can be ALD coated on Ag nanowires to enhance the thermal stability [58].

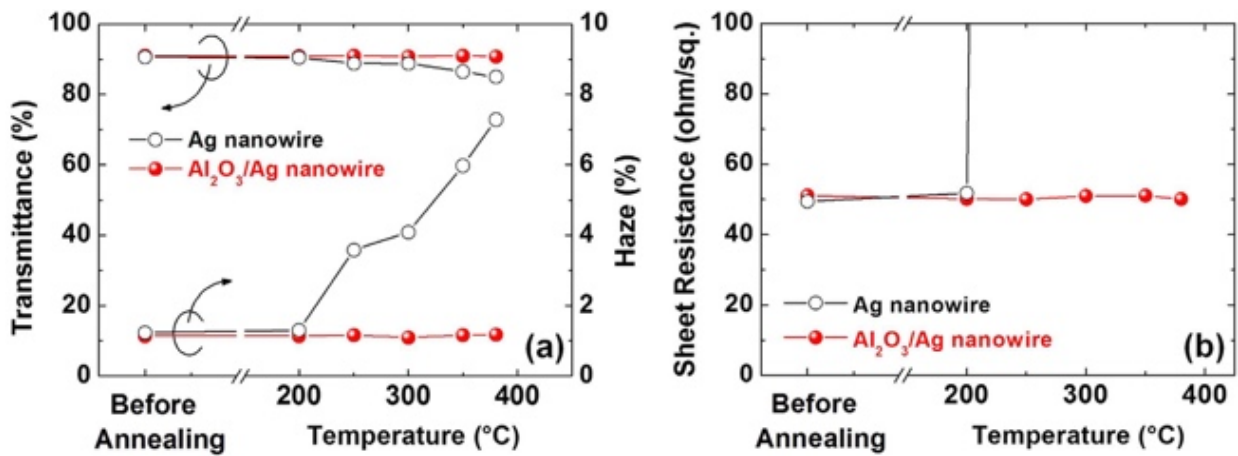

**Fig. 8.** Changes in (**a**) optical transmittance/haze and (**b**) sheet resistance of the Ag and $Al_2O_3$/Ag nanowire electrodes as a function of the annealing temperature. The annealing time is 20 min. Reprinted from (Hwang, B.; An, Y.; Lee, H.; Lee, E.; Becker, S.; Kim, Y.-H.; Kim, H. *Sci. Rep.* **2017**, *7*, 41336. DOI:10.1038/srep41336.).

### 3.4. Tuning of the optical properties

The total transparency of a TE includes the transmittance, haze and clarity. Transmittance refers to the amount of light that passes through the TE without being scattered. Haze and clarity are the amounts of light that are subject to wide and narrow scattering, respectively. Most studies of TEs focus on the transmittance and the haze. In some applications, high transmittances and low haze are desired, such as in touch panels [12], while in some applications, a high haze is desired, such as in solar cells [59].

The haze of Ag nanowire films depends on the diameter and length of the nanowires. For a single nanowire, the diameter determines the haze. However, for a nanowire film, the lengths of the nanowires are of great importance because the lengths determine the density of the film (also called the area fraction [60] of the nanowires). Therefore, the optical haze can be reduced by using longer and thinner nanowires [61].

Based on the above, it seems straightforward to make a TE with high transmittance and low haze by using thin nanowires [62]. However, TEs with thin nanowires have lower thermal and chemical stability compared to those with thick nanowires, thereby shortening their lifetime. This requires additional coatings. One route to obtain low haze TEs is to use ultra-long Ag nanowires with relatively thick diameters [63]. Another route to reduce the haze is to coat the Ag nanowires with materials that have smaller extinction coefficients (Fig. 9), such as gold [64].

It also seems reasonable to consider making a high haze Ag TE using short and thick nanowires. The problem in this case is that it reduces the transmittance. In practice, researchers use extra coatings, such as FZO, on films made of thin nanowires to increase the optical haze [9].

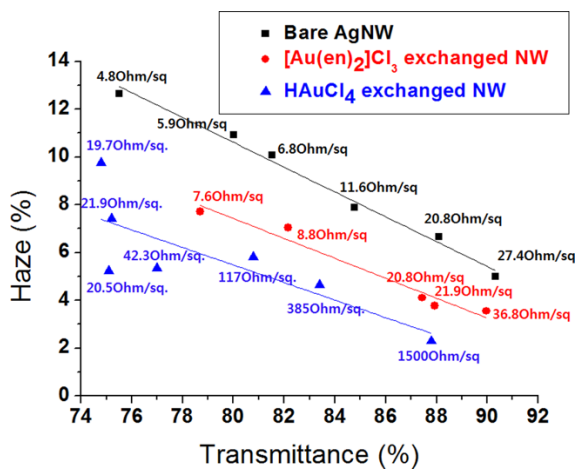

**Fig. 9.** Comparison of haze and transmittance for transparent films fabricated with bare Ag nanowires, Au-coated Ag nanowires made using HAuCl$_4$ exchange, and Au-coated Ag nanowires made using [Au(en)$_2$]Cl$_3$ exchange and NH$_3$ treatment. The sheet resistance for each sample is also stated. Reprinted with permission from (Kim, T.; Canlier, A.; Cho, C.; Rozyyev, V.; Lee, J. Y.; Han, S. M. *ACS Appl. Mater. Interfaces* **2014**, *6* (16), 13527–13534. DOI:10.1021/am502632t). Copyright (2014) American Chemical Society.

## Conclusion and future perspectives

Silver nanowire-based transparent electrodes (TEs) are promising candidates for replacing transparent conductive oxide (TCO) electrodes. Many studies have shown their excellent mechanical, electrical, optical, and chemical properties, enabling new applications that are not achievable using TCO electrodes. Many protocols have been developed for fabricating Ag nanowire-based TEs with the aim of industrial production.

It is generally recognized that the requirements of industrial production are a transmittance > 90% and a sheet resistance < 100 Ω/sq. However, these are two basic requirements, and in practice, more requirements must be fulfilled for production. Fig. 10 shows the five basic factors for evaluating transparent electrodes: electrical, optical, mechanical, chemical and thermal properties. Each of these factors contains one or more elements that must also be evaluated.

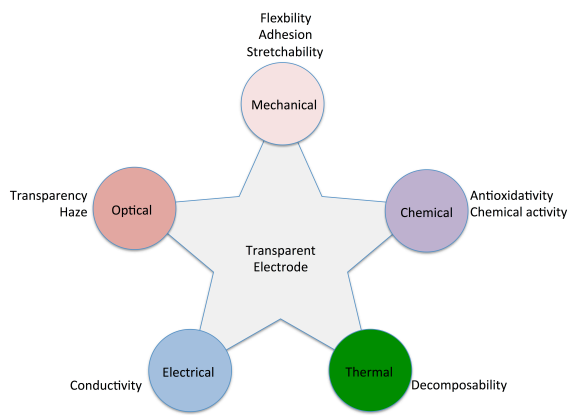

**Fig. 10.** Schematic drawing of the basic factors for evaluating transparent electrodes.

**Acknowledgments**

The authors thank the Knowledge Foundation, the Swedish Energy Agency, the EU Regional Funds and the County Administrative Board in Västernorrland, Sweden, for financial support.